\documentstyle[twoside,fleqn,espcrc2,epsfig]{article}
\graphicspath{{/l3/guide/l3pictures/detector/all/},
              {/hp/gen1a/user/braccini/leap98/proceedings/fig/}}
\def\ra{\rightarrow }

\def\epem{\mbox{e}^+\mbox{e}^- }
\def\pip{\pi^+ }
\def\pim{\pi^- }

\def\kos{\mbox{K}^0_S }
\def\k{\mbox{K}}      
\def\pb{pb$^{-1}$}
\def\NP{{\it Nucl. Phys. }}
\def\PL{{\it Phys. Lett. }}
\def\ZfP{{\it Z. Phys. }}
\def\NIM{{\it Nucl. Inst. Meth. }}
\def\PRep{{\it Phys. Rep. }}
\def\PR{{\it Phys. Rev. }}
\def\PRL{{\it Phys. Rev. Lett. }}

\newcommand{\AmS}{{\protect\the\textfont2
  A\kern-.1667em\lower.5ex\hbox{M}\kern-.125emS}}

\title{The K$^0_s$K$^0_s$ Final State in Two-Photon Collisions
and Some Implications for Glueball Searches}

\author{Saverio Braccini\address{University of Geneva - DPNC\\
        24, Quai Ernest Ansermet \\
        CH-1211 Gen\`eve 4, Switzerland\\
	E-mail: Saverio.Braccini@cern.ch}%
        \thanks{Invited contribution to LEAP 98, Villasimius (Italy), 
	September 1998.}}
       
\begin{document}
%
%
%
\begin{abstract}
The K$^0_s$K$^0_s$ final state in two-photon collisions is studied with
the L3 detector at LEP using data at centre of mass energies from 91 GeV
to 183 GeV. The K$^0_s$K$^0_s$ mass spectrum is dominated
by the formation of the f$_2\,\!\!\!'$(1525) tensor meson whose two-photon partial
width is measured. Clear evidence for destructive f$_2$-a$_2$ interference is observed.
No signal is present in the region around 2.2 GeV. An upper limit for
the two-photon partial width times the  K$^0_s$K$^0_s$ branching ratio of the $\xi$(2230) glueball candidate is
then derived. An enhancement is observed around 1750 MeV. It may be due to the formation of
a radial recurrence of the f$_2\,\!\!\!'$(1525) or to the $s\bar{s}$ member of the $0^{++}$ meson nonet.
\end{abstract}

\maketitle

\section{Introduction}
\label{intro}
Electron-positron storage rings are widely used to investigate 
the behaviour of two-photon interactions via the process
$\epem\ra\epem\gamma^*\gamma^*\ra\epem \mbox{X}$, where $\gamma^*$ is a 
virtual photon. 
The outgoing electron and  positron carry nearly  the full 
beam energy and their transverse momenta are usually so small that
they are not detected. 
This kind of event is 
characterised by an initial state $\epem\gamma^*\gamma^*$, calculable by QED, and
a low multiplicity final state. This process is
particularly useful in the study  of the
properties of hadron resonances.

The total cross section $\sigma_{T}$ for the formation of a resonance R is given by:
\begin{equation}
\sigma_{T}(\epem\ra\epem\mbox{R})=\int d^5{\cal{L}}_{\gamma\gamma}
\sigma(\gamma^*\gamma^*\ra\mbox{R})
\label{eq:stot}  
\end{equation}
where $d^5\cal{L}_{\gamma\gamma}$ is the differential luminosity
function giving the flux of virtual photons.
For quasi-real photons $\sigma(\gamma^*\gamma^*\ra$R) is given by
the Breit-Wigner formula:
\begin{equation}
\sigma
= 8 \pi (2J_R+1) 
\frac{\Gamma_{\gamma\gamma}(\mbox{R})\Gamma(\mbox{R})}
{(W_{\gamma \gamma}^2-m_R^2)^2+m_R^2\Gamma^2(\mbox{R})}
\label{eq:sgg}  
\end{equation}
where $W_{\gamma\gamma}$ is the invariant mass of the two-photon
system, $m_R$, $J_R$, $\Gamma_{\gamma\gamma}(\mbox{R})$
and $\Gamma($R)
are the
mass, spin, two-photon partial width 
and total width of the resonance, respectively.
Combining equations~(\ref{eq:stot}) and~(\ref{eq:sgg}) 
leads to the proportionality
relation  
\begin{equation}
\sigma_{T}(\epem\ra\epem\mbox{R})={\cal{K}}\cdot\Gamma_{\gamma \gamma}(\mbox{R})
\label{eq:prop}  
\end{equation}
where the
proportionality factor ${\cal{K}}$ can be evaluated by a
Monte Carlo integration. 
Equation~(\ref{eq:prop}) 
is used to determine
the two-photon partial width of the resonance.

The quantum numbers of the 
resonance must be compatible with the initial state of the two quasi-real
photons. A neutral, unflavoured meson with even charge conjugation
and helicity-zero or two can be formed. In order to decay into $\kos\kos$, 
the resonance must have
$J^{PC}=(\mbox{even})^{++}$. 

For the $2^{++}$, 1$^3$P$_2$ tensor meson  nonet, the f$_2$(1270), the a$_2^0$(1320) 
and the  f$_2\,\!\!\!'$(1525) can be formed.
However, since these three states are close in mass,
interferences must be taken into account.
According to SU(3), 
the f$_2$(1270) interferes constructively with the a$_2^0$(1320) in the
$\mbox{K}^+\mbox{K}^-$ final state but destructively 
in the $\k^0\bar{\k^0}$ final state~\cite{Lipkin1}.
Therefore, among the states of the tensor meson nonet,
only the f$_2\,\!\!\!'$(1525) was observed by previous 
experiments in the $\kos\kos$ final state~\cite{Cello}.

Since gluons do not couple to photons, a pure glueball couples to two photons
only via a box diagram and its two photon width is therefore expected to be
very small. A state that can be formed in a gluon rich environment but
not in two photon fusion has the typical signature of a glueball.

An analysis of the reaction $\epem\ra\epem\kos\kos$ is presented here, 
where only the $\kos\ra\pip\pim$ decay is considered. The data correspond
to an integrated luminosity of 143 \pb collected by the L3 detector at LEP 
at $\sqrt s=91$ GeV and 52 \pb at $\sqrt s=183$ GeV. 

A study of the $\kos\kos$ final state in two-photon collisions was already
published by the L3 Collaboration~\cite{Saverio} with a luminosity of 114 \pb 
at $\sqrt s=91$ GeV.

The L3 experiment is described in detail elsewhere~\cite{l3const}.
In this analysis, the charged particle tracker is mainly used.
It is composed by a silicon microstrip vertex detector and
a multiwire drift chamber. The events are triggered by a 
low $p_t$ threshold charged-track trigger~\cite{trigger}.

\section{ Event Analysis}

\begin{figure}[t]
\begin{center}
  \mbox{\epsfig{file=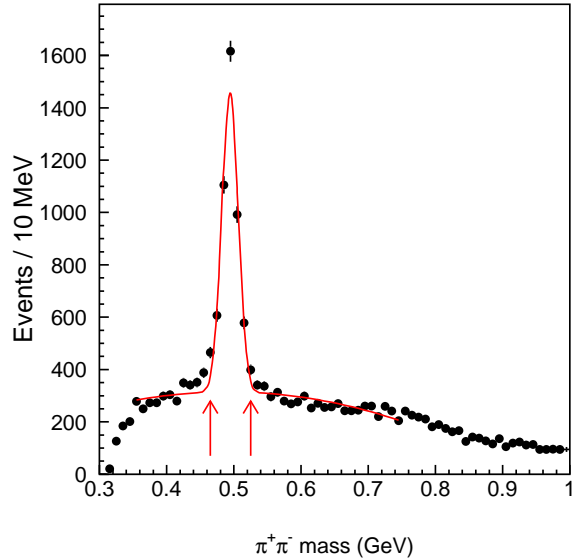,width=7.5cm}}
\end{center}
\vspace*{-1.3cm}
\caption {The $\pip\pim$  mass spectrum
for reconstructed secondary vertices which are 
more than 3 mm from the interaction point.
The curve is the result of a fit using a third-order polynomial 
for the background and a Gaussian 
for the peak. The arrows indicate the signal region. }
  \label{fig:k}
\end{figure}

In order to select $\epem\ra \epem\pip\pim\pip\pim$ 
events, we require:
\begin{itemize}
\item The total energy seen in the calorimeters must be smaller than 30 GeV
to exclude annihilation events. 
\item There must be exactly four good charged tracks in the tracking chamber with a net
charge of zero. A good track requires more than 20 hits out of a maximum of 62. 
\item The total momentum 
imbalance in the transverse plane must satisfy:\\
$|\sum \overrightarrow{p_T}|^2 < 0.1\hspace{.2cm} \mbox{GeV}^2$.
\item Events with photons are rejected.
A photon is defined as an isolated shower in the electromagnetic calorimeter
with an energy larger than 100 MeV. 
\end{itemize}

\begin{figure}[t]
  \begin{center}
  \mbox{\epsfig{file=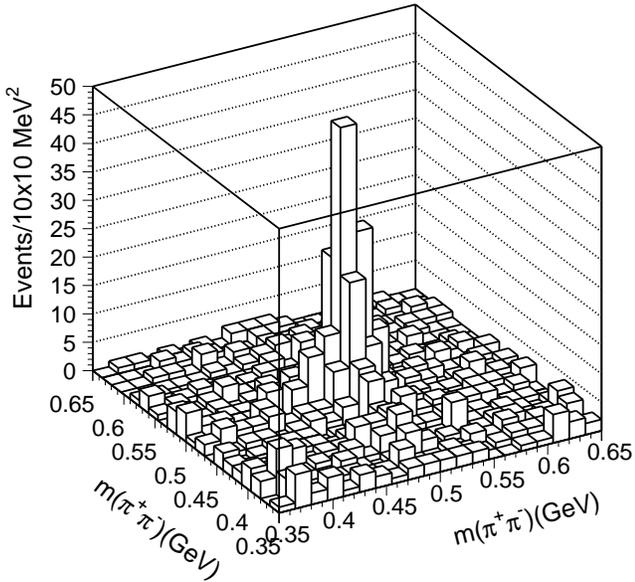,width=7.cm}}
\end{center}
\vspace*{-.7cm}
\caption {
 m($\pi^+\pi^-$) at one  secondary vertex versus m($\pi^+\pi^-$) at the other 
secondary vertex. There is a strong enhancement near the $\kos\kos$ point over
a very small background. }
  \label{fig:kk}
\end{figure}

The $\kos$'s are identified by requiring a secondary vertex distinct
from the primary interaction point with a distance greater than 1 mm in
the transverse plane. 
In order to select $\kos\kos$ exclusive events, we require:
\begin{itemize}
\item At least one of the two secondary vertices must be at
a distance greater than 3 mm from 
the interaction point in the transverse plane.
\item The angle between the flight direction of each $\kos$ candidate
(taken as the line between the interaction point and the secondary vertex
in the transverse plane)
and the total transverse momentum vector of the
two outgoing tracks must be less than 0.3 rad.
\item Since the two $\kos$'s are produced back-to-back 
in the transverse plane, the angle between
the flight directions of the two $\kos$ candidates
in this plane is required to be $\pi\pm 0.3$ rad. 
\end{itemize}

\begin{figure}[t]
  \begin{center}
  \vspace*{-.8cm}
      \mbox{\epsfig{file=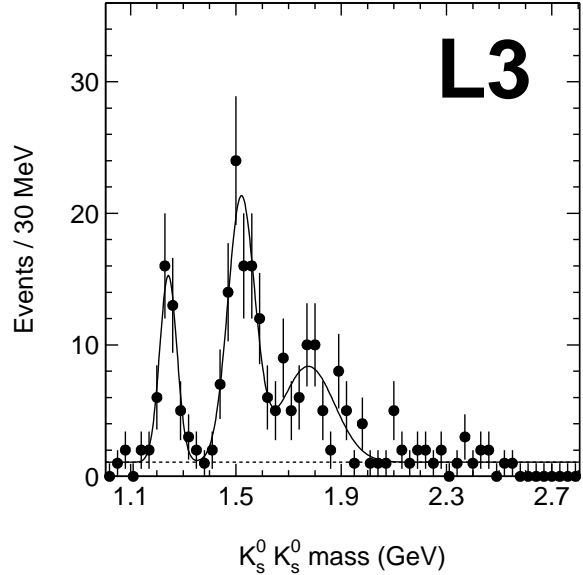,height=8cm}}\\
  \end{center}
\vspace*{-1.3cm}
\caption{The $\kos\kos$ invariant mass spectrum: the solid line
corresponds to the maximum likelihood fit. The background is
fitted by a constant (dashed line) and the three peaks by 
Gaussian curves. }  
  \label{fig:kk_sp}
\end{figure}

Fig.~\ref{fig:k} shows
the $\pip\pim$ mass distribution with a 3 mm  vertex cut for the $\kos$. 
Fitting this distribution, a mass resolution
of $\sigma=9.6\pm0.8$ MeV is found, consistent with the Monte Carlo simulation.   
The invariant masses of the two $\kos$ candidates must be inside
a circle of 40 MeV radius around the $\kos\kos$ point (Fig.~\ref{fig:kk}).

With these selection criteria, 253 events are found in the data sample.
The background due to misidentified $\kos$ pairs is estimated to be negligible
by a study of the $\kos$ sidebands.  
The  background  due to the $\kos\k^\pm \pi^\mp$ final state 
is determined to be negligible by a  Monte Carlo simulation. 
The beam-gas and beam-wall contributions are found negligible.

The resulting $\kos\kos$ invariant mass
spectrum is shown in 
Fig.~\ref{fig:kk_sp}. The spectrum is dominated by the f$_2\,\!\!\!'$(1525) resonance.
The f$_2(1270)-$a$_2^0(1320)$ region shows the destructive f$_2-$a$_2^0$ interference in  the
$\kos\kos$ final state~\cite{Lipkin1}. A clear enhancement is visible in the
1750 MeV region. No excess is present around 2230 MeV.

A maximum likelihood fit using three Gaussians plus a constant is 
then performed on the $\kos\kos$ mass spectrum. The masses and widths of
the Gaussians are left free in the fit.
The fit is shown in Fig.~\ref{fig:kk_sp} and the results are summarised in 
Table~\ref{tab:li_fit}.

\begin{table}[t]
\caption{Results from the maximum likelihood fit of the $\kos\kos$ 
invariant mass spectrum.}
\begin{center}
  \begin{tabular}{ccc}\hline
\rule[.15in]{0.0in}{0.0in} & f$_2\,\!\!\!'$(1525)  & 1750 MeV Region\\ \hline
Mass  (MeV)      & 1520 $\pm$  7    & 1770 $\pm$ 20   \\ 
Width (MeV)      &  119 $\pm$ 15    &  200 $\pm$ 50   \\ 
Nb. of events &   85 $\pm$ 11    &   61 $\pm$ 10  \\ \hline
  \end{tabular}
  \vspace*{-1.cm}
  \label{tab:li_fit}
 \end{center}
\end{table}

The f$_2\,\!\!\!'$(1525) statistical significance 
is about 8 standard deviations,
the statistical significance of the enhancement around 1750 MeV   
is about 6 standard deviations.
From the fit 38$\pm$7 events are found in the f$_2$-a$_2$ region.

In order to correct the data for the detector acceptance and efficiency, 
a Monte Carlo procedure is used~\cite{Linde}. 
The nominal f$_2\,\!\!\!'$(1525) parameters~\cite{PDG} 
are used for the generation. 
The angular distribution of the two $\kos$'s in the two-photon center-of-mass system
is generated according to
phase space i.e. uniform in $\cos\theta^*$ and in $\phi^*$, where
$\theta^*$ and $\phi^*$ are the polar and azimuthal angles taking
the $z$ direction parallel to the electron beam.
In order to take into account the helicity of a spin-two resonance, 
a weight is assigned to each generated event according to the 
weight functions:  
$w=(\cos^2\theta^*-\frac{1}{3})^2$ for the helicity-zero contribution and
$w=\sin^4\theta^*$ for the helicity-two contribution.
 
All the events are passed through the full detector simulation program 
and are reconstructed following the same procedure used for the data.
Although the detector acceptance is rather high (15\% for helicity-zero
and 30\% for helicity-two), the trigger efficiency (83\%)
and the analysis cuts ($\sim 30\%$)
give a total efficiency of 3.9\% for helicity-zero 
and 7.6\% for helicity-two for the f$_2\,\!\!\!'$(1525) at $\sqrt s=91$ GeV.
At $\sqrt s=183$ GeV the total efficiencies are 3.1\% and 7.2\% respectively.

\section{Results}

\begin{figure}[t]
  \begin{center}
      \mbox{\epsfig{file=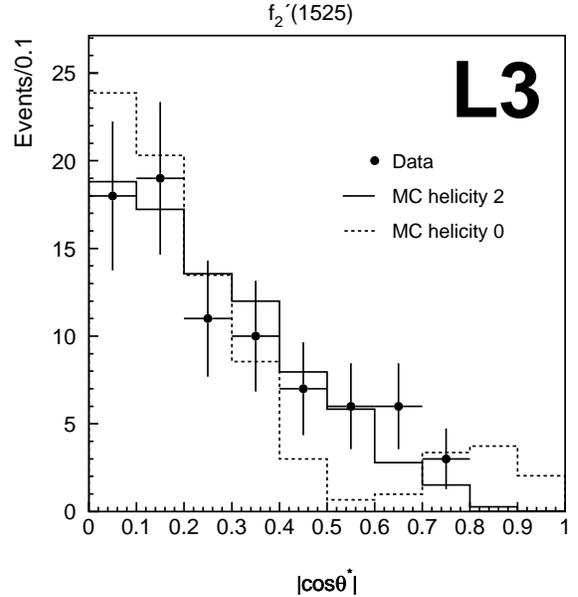,width=7.cm}}\\
  \end{center}
\vspace*{-.8cm}
\caption{The $\kos\kos$ polar angular distribution 
compared with the Monte Carlo distributions for the hypothesis 
of a pure helicity-zero and helicity-two 
contribution for the f$_2\,\!\!\!'$(1525). 
The Monte Carlo curves are normalised to the same number of events 
as the data. }
  \label{fig:ad_f2}
\end{figure}

\subsection{The f$_2\,\!\!\!'$(1525) tensor meson}
In order to determine the helicity state, 
a study of the angular distribution of the two $\kos$'s
from f$_2\,\!\!\!'$ decay in the
two-photon center of mass is performed.
The experimental polar angle distribution is compared with the
Monte Carlo  in 
Fig.~\ref{fig:ad_f2} for both the helicity-zero and helicity-two cases.
The Monte Carlo distributions are normalised to the same number of
events as in the data and no background subtraction is done.
The $\chi^2$ values for the helicity-zero and helicity-two hypotheses 
are 81 and 7 for nine
degrees of freedom respectively. 
Thus pure helicity-zero is
excluded and helicity-two is largely dominant, 
in agreement with the theoretical predictions~\cite{Kopp}.

The total cross section $\sigma_T$ times the branching ratio Br into $\k\bar{\k}$ for the f$_2\,\!\!\!'$(1525)
is measured using the formula
\begin{equation}
\sigma_T\times\mbox{Br} 
=\frac{N_{obs}-N_{back}}{{\cal{L}}\varepsilon}
\end{equation}
where $\cal{L}$ is the integrated luminosity
and $\varepsilon$ is the 
total efficiency. The number of signal events 
$N_{obs}-N_{back}$ is determined from the
maximum likelihood fit. From our data only 
$\sigma_{T}\times $Br(f$_2\,\!\!\!'\ra\kos\kos\ra\pip\pim\pip\pim$) can be measured.
Using the PDG~\cite{PDG}
value for Br($\kos\ra\pip\pim$) and 
Br(f$_2\,\!\!\!'\ra\k\bar{\k})=4\times$Br(f$_2\,\!\!\!'\ra\kos\kos$) from
isospin symmerty, 
$\sigma_{T}\times$Br(f$_2\,\!\!\!'\ra$ K$\bar{\mbox{K}}$)
can be determined. 

The product $\Gamma_{\gamma\gamma}$(f$_2\,\!\!\!')\times$ Br(f$_2\,\!\!\!'\ra\k\bar{\k})$
is measured from the cross section using the formula
\begin{equation}
\Gamma_{\gamma\gamma}(\mbox{f}_2\,\!\!\!')\times\mbox{Br(f}_2\,\!\!\!'\ra\k\bar{\k})=
\frac{\sigma_T\times\mbox{Br(f}_2\,\!\!\!'\ra\k\bar{\k})}{{\cal{K}}}
\end{equation}
where the proportionality factor ${\cal{K}}$ is evaluated by 
Monte Carlo integration. Two separate measurements are performed for data collected at $\sqrt s=91$ GeV and
$\sqrt s=183$ GeV.
The results are in very good agreement with the published value~\cite{Saverio}
$\Gamma_{\gamma\gamma}$(f$_2\,\!\!\!')\times $Br(f$_2\,\!\!\!'\ra\k\bar{\k})=(0.093\pm 0.018\pm 0.022) \mbox{keV}$
under the hypothesis of a pure helicity-two state.

%
%
\subsection{The 1750 MeV region}

\begin{figure}[t]
  \begin{center}
      \mbox{\epsfig{file=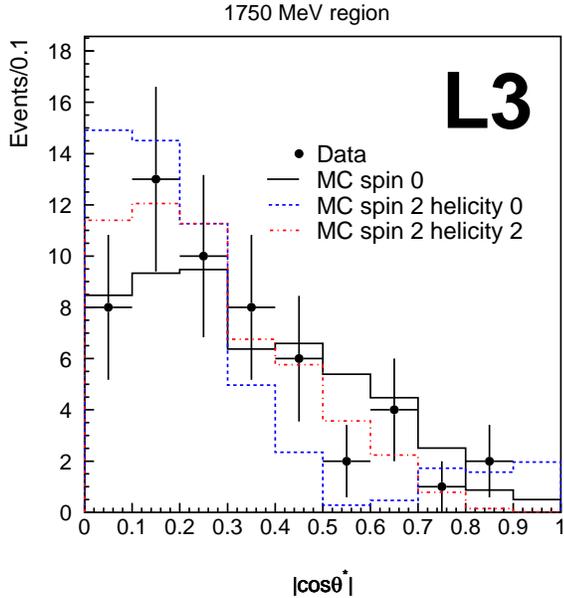,width=7.cm}}\\
  \end{center}
\vspace*{-.8cm}
\caption{The $\kos\kos$ polar angular distribution 
compared with the Monte Carlo distributions for the hypothesis 
of a pure spin two helicity-zero, spin two helicity-two and spin zero 
contribution in the 1750 MeV region. 
The Monte Carlo curves are normalised to the same number of events 
as the data.}
  \label{fig:ad_18}
\end{figure}

The enhancement around 1750 MeV may be due to the formation of a radially
excited f$_2'$ state, according to theoretical predictions~\cite{Munz}.
However, the 1750 MeV region is very interesting especially looking at the recent results
by the Crystal Barrel Collaboration~\cite{CryBar}. They report the observation of three
$J^{PC}$=0$^{++}$ states the a$_0$(1450), the f$_0$(1370) and the f$_0$(1500). 
In the scenario presented in~\cite{Close} the a$_0$(1450)
is the isovector member of the 0$^{++}$ meson nonet while the f$_0$(1370) 
and the f$_0$(1500) cannot both be
isoscalar members of the nonet. The f$_0$(1370) is very likely 
mainly composed by u and d quarks while the f$_0$(1500) is compatible with the ground state
scalar glueball expected around 1500 MeV. This hypothesis includes the prediction of
a further scalar state, the f$'_0$(1500-1800), mainly composed by s quarks. This state
will couple strongly to K$\bar{\mbox{K}}$ and its discovery is essential to support
the f$_0$(1500) glueball nature. 

In order to investigate the spin of the 1750 MeV region, 
the angular distribution of the two $\kos$'s
in the two-photon center of mass is studied.
The polar angle distribution is compared in Fig.~\ref{fig:ad_18} 
with the Monte Carlo predictions
for the spin two helicity-zero (J=2, $\lambda$=0), spin two helicity-two (J=2, $\lambda$=2) and spin zero 
(J=0) cases.
The Monte Carlo distributions are normalised to the same number of
events as in the data and no background subtraction is done.
The $\chi^2$ values are 50, 26 and 7 for nine degrees of freedom
for J=2 $\lambda$=0, J=2 $\lambda$=2 and J=0
hypotheses respectively.
J=0 assignment is found 3.2 times more probable than J=2 $\lambda$=2.

Experimental indications for a scalar state around 1750 MeV decaying into
K$^+$K$^-$~\cite{BESKPKM} and $\eta\eta$~\cite{BUGG} have been recently found.

%
%
\subsection{The 2230 MeV region}
The BES Collaboration confirmed the previous observation by the Mark III Collaboration
of a resonance, 
the $\xi$(2230)~\cite{xi2230}, produced in radiative decays of the J/$\psi$ particle. 
Due to its narrow width and its production in gluon rich environment,
this state is considered a glueball candidate.
Its mass is consistent with the lattice QCD prediction for the
ground state tensor glueball. 

Since gluons do not couple to photons, the two-photon width
is expected to be small for a glueball. To make this statement more quantitative, a parameter called
stickiness~\cite{Chanowitz} can be introduced for a state X:
\begin{eqnarray}
S_X&=&N_l \left(\frac{m_X}{k_{J/\psi\rightarrow\gamma X}}\right) ^{2l+1} 
\frac{\Gamma(J/\psi\rightarrow\gamma X)}{\Gamma(X\rightarrow\gamma\gamma)}\\
&\sim&\frac{|<X|gg>|^2}{|<X|\gamma\gamma>|^2} \nonumber
\end{eqnarray}  
where $m_X$ is the mass of the state, $k_{J/\psi\rightarrow\gamma X}$ is the energy of the photon from a
radiative J/$\psi$ decay in the J/$\psi$ rest frame and $l$ is the angular momentum between the two gluons.
For spin two states $l=0$. $N_l$ is a normalisation factor that can be calculated assuming by
definition the stickiness of the f$_2$(1270) tensor meson to be 1. 

Using the same method adopted for the f$_2\,\!\!\!'$(1525), a Monte Carlo simulation is used to
determine the detection efficiency for the $\xi$(2230). For the simulation we use values for the mass and the width
determined by combining the results by Mark III and BES. A mass resolution
of $\sigma$= 60 MeV is found by fitting a Gaussian to the $\kos\kos$ Monte Carlo spectrum.
 The total detection efficiency is measured to be 16.5\% at $\sqrt s=$91 GeV
and 9.3\% at  $\sqrt s=$183 GeV
under the hypothesis of a pure helicity-two contribution.
The signal region is chosen to be $\pm2\sigma$ around the 
$\xi$(2230) mass. In order to evaluate the background two sidebands of $2\sigma$ are considered.
At $\sqrt s=$91 GeV and at $\sqrt s=$183 GeV,
6 and 2 events are found in the signal region respectively. Fitting a constant in the sideband region, the expected
background are evaluated to be 4.9 and 7.9 events. Using the standard method~\cite{PDG} for extracting 
an upper limit for a Poisson distribution with background, we determine upper limits of 7.3 and 3.6 events respectively
at 95\% C.L. Using the same method adopted for measuring the two-photon width of the f$_2\,\!\!\!'$(1525),
an upper limit of $\Gamma_{\gamma\gamma}(\xi(2230))\times $Br$(\xi(2230)\ra\kos\kos)$ is derived combining the
results from the two values  of $\sqrt s$. The combination is performed using
a Monte Carlo technique.
We obtain $\Gamma_{\gamma\gamma}(\xi(2230))\times $Br$(\xi(2230)\ra\kos\kos)<3$ eV 
at 95\% C.L. under the hypothesis of a pure helicity-two state.

Combining the results reported by BES and Mark III for the $\Gamma(J/\psi\rightarrow\gamma\xi)\times $Br$(\xi\ra\kos\kos)$
and our upper limit on $\Gamma(\xi\rightarrow\gamma\gamma)\times $Br$(\xi\ra\kos\kos)$, we obtain
a lower limit on the stickiness $S_{\xi(2230)} >$33 at 95\% C.L. 
Errors are taken into account by a Monte Carlo calculation.
This value is much larger than
the values measured for all the established $q\bar{q}$ states.

\section{Conclusions}
The reaction 
$\epem\ra \epem\gamma^*\gamma^*\ra\epem\kos\kos$ 
is studied with the L3 detector at LEP.
The spectrum is dominated by the f$'_2$(1525) tensor meson.
The angular distribution excludes pure helicity-zero and
shows that helicity-two is largely dominant.
The $\Gamma_{\gamma\gamma}(\mbox{f}_2\,\!\!\!')\times\mbox{Br}(\mbox{f}_2\,\!\!\!'\ra
\k\bar{\k}) $ is found to be consistent with the previously published value~\cite{Saverio}.
Clear evidence for destructive f$_2$-a$_2$ interference is observed.
An enhancement of about 6 standard deviations is observed around 1750 MeV. 
It may be due to the formation of a
radial recurrence of the f$'_2$(1525) tensor meson (J=2, $\lambda=$2)
but the observed angular distribution favours the J=0 assignment. Therefore the enhancement 
may be due
to the formation of the $s\bar{s}$ member of the $0^{++}$ meson nonet.
The latter case strongly supports the glueball nature of the f$_0$(1500)~\cite{Close}.
No signal is present in the region of the $\xi$(2230) glueball candidate. 
We obtain the upper limit
$\Gamma_{\gamma\gamma}(\xi(2230))\times $Br$(\xi(2230)\ra\kos\kos)<3$ eV 
at 95\% C.L. under the hypothesis of a pure helicity-two state.

\section*{Acknowledgments}
I would like to express my gratitude to the two-photon physics group
of the L3 collaboration in particular to M.N. Focacci-Kienzle,
J.H. Field and B. Monteleoni.

%

\end{document}